\documentstyle[a4wide,epsf]{article}
\newcommand{\rcite}[1]{{\cite{#1}}}
\newcommand{\rref}[1]{{(\ref{#1})}}
\newcommand{\tref}[1]{{\ref{#1}}}
\newcommand{\rlabel}[1]{{\label{#1}}}
\newcommand{\rbibitem}[1]{\bibitem{#1}}
\newcommand{\be}{\begin{equation}}
\newcommand{\ee}{\end{equation}}
\newcommand{\ba}{\begin{eqnarray}}
\newcommand{\ea}{\end{eqnarray}}

\newcommand{\trace}[1]{\langle #1 \rangle}
\newcommand{\Vpar}{{V_\parallel}}
\newcommand{\Vper}{{V_\perp}}
\begin{document}
\begin{titlepage}
\begin{flushright}
LU TP 97/03 \\
NORDITA-97/22 N/P \\
hep-ph/9704212\\ 
March 1997 (revised July 1997)
\end{flushright}

\vfill

\begin{center}
\begin{bf}
{\Large \bf Vector Meson Masses in Chiral Perturbation Theory} \\[2cm]
\end{bf}
J. Bijnens$^a$, P. Gosdzinsky$^b$ and P. Talavera$^a$ \\[1cm]
$^a$Department of Theoretical Physics, University of Lund \\
S\"olvegatan 14A, S-22362 Lund, Sweden. \\[0.5cm]
$^b$NORDITA, Blegdamsvej 17, DK-2100 Copenhagen \O , Denmark. \\ 

\vfill

{\bf PACS:} 12.39.Fe, 12.40.Yx, 12.15.Ff, 14.40.-n
\end{center}
\vfill
\begin{abstract}
We discuss the vector meson masses within the context of Chiral Perturbation
Theory performing an expansion in terms of the momenta,
quark masses and $1/N_c$. 
We extend the previous analysis to include isospin breaking effects
and also include up to order $p^4$. We discuss vector meson chiral
perturbation theory in some detail and present a derivation from
a relativistic lagrangian. The unknown coefficients are estimated
in various ways. We also discuss the relevance of electromagnetic
corrections and the implications of the present calculation for the
determination of quark masses.
\end{abstract}
\vfill
\end{titlepage}

\section{Introduction}
\rlabel{intro}

The relevance of vector meson masses and mixings in the determination
of the quark masses is well known. A review is in \rcite{GL0}.
Virtually all the information used in that paper was at the lowest order in
Chiral Perturbation Theory and at most the leading nonanalytic corrections are
included. The main parts of the information used for quark mass ratios were
pseudoscalar, baryon and vector meson masses and mixings.
The full one-loop corrections to the pseudoscalar meson masses and
mixings was performed in \rcite{GL1} and a recent review on their
relevance to quark mass ratios is in \rcite{Leutwyler}.
The subject of baryon masses in chiral perturbation theory is a subject
of a lot of current research. For recent results see \rcite{baryons} and
references therein. This paper is devoted to a similar study for the vector
meson masses. 

Chiral Perturbation Theory for Vector Mesons using a heavy meson
formalism was first introduced in \rcite{ELISABETH}. We will refer to this as
the Heavy Meson Effective Theory (HMET). There the first correction
of order $p^3$ was evaluated. This was later extended to include isospin
breaking effects in \rcite{PETERHANS}. In that reference an estimate
of the electromagnetic effects using a short-long-distance matching
calculation was also performed. The result for the quark masses from
$\rho^0-\omega$ mixing agreed well with standard expectations, but the
$K^{*0}$--$K^{*+}$ mass difference did not. The size of the chiral
corrections of order $p^3$ is rather large so a more complete calculation
seemed necessary. We therefore perform in this paper the full vector
meson mass and mixing corrections to order $p^4$.
We also discuss the electromagnetic corrections.

The paper is organized as follows. In Sect. \tref{basic} we give the main
notation and conventions. In the next section we explain the HMET
and list all of the terms necessary for the calculation of the
vector meson masses. Then in Sect. \tref{relativistic} we explain
the connection between the relativistic descriptions and the HMET.
In Sect. \tref{coefficients} we then give the model estimates for the
various parameters in the HMET, followed in Sect. \tref{calculation} by a
description of the calculation of the main result
in this paper, the vector meson
masses to order $p^4$. We then discuss how we include the electromagnetic
correction in Sect. \tref{electromagnetic} and the numerical results
in Sect. \tref{numerics}. We then give the main conclusions and the
implications for ratios of quark masses in Sect. \tref{conclusions}.

In App. \tref{integrals} we quote the integrals we use in all relevant
kinematical domains and in App. \tref{formulas} we give the
approximate analytical formulas for the vector masses, as described in
Sect. \tref{calculation}.

\section{Basic Ingredients}

\rlabel{basic}
In this section we put some of the basic notation.
The notation is essentially the same as in \rcite{TONI} and \rcite{EDUARD}.

The pseudo-Goldstone boson fields can be written as a 
$3\otimes 3$ special
unitary matrix 
\be 
U = \exp {i\sqrt{2}{\Pi}\over F} = u^2\rlabel{eq:1}
\ee
where
\be
{\Pi} = \left[
\begin{array}{ccc}
{\pi^0\over\sqrt{2}} + {\eta\over \sqrt{6}} & \pi^+ & K^+ \\ \pi^- & -{\pi^0
\over \sqrt{2}} + {\eta \over \sqrt{6}} & K^0 \\ K^- & \overline K^0 & - {2\eta
\over \sqrt{6}}
\end{array} \right]
\ee
and $F \sim F_\pi = 92.4$ MeV.\\
The relativistic vector meson fields are introduced as a $3 \otimes 3$ nonet matrix
(in the $N_C$ limit)
\be
 V_\mu = \left[
\begin{array}{ccc}
{\rho^0_\mu \over \sqrt{2}} + {\omega\mu \over \sqrt{2}} & \rho_\mu^+ &
K_\mu^{*+} \\ \rho_\mu^- & -{\rho_\mu^0 \over \sqrt{2}} + {\omega\mu \over
\sqrt{2}} & K_\mu^{*0} \\ K_\mu^{*-} & \overline K_\mu^{*0} & {\phi_\mu
}
\end{array} \right],
\rlabel{vectorMATRIX}
\ee
In what follows
$\langle A \rangle$ denotes trace of $A$.
We also use the quantities
\ba
\chi_+ &=& u^\dagger \chi u^\dagger+u  \chi^\dagger u 
\nonumber \\
\chi &=& 2 B_0(s+ip)
\nonumber \\
u_\mu &=&  i u^\dagger D_\mu U u^\dagger =i(u \partial^\mu u^\dagger-u^\dagger
\partial^\mu u) 
\nonumber \\
V_{\mu \nu} &=& D_\mu V_\nu - D_\nu V_\mu \nonumber \\
D_\mu V_\nu& =& \partial_\mu V_\nu+[\Gamma_\mu,V_\nu] \nonumber \\
\Gamma_\mu&=&\frac{1}{2}(u \partial_\mu u^\dagger+u^\dagger
\partial_\mu u) \rlabel{1DEFS}
\ea
In \rref{1DEFS}, we have ignored the vector and axial external sources.
We will also set the pseudoscalar source, $p$, equal to
zero and $s$ will be the diagonal matrix
$(m_u,m_d,m_s)$. The parameter $B_0$ is given by
\be
\langle 0 | \bar q q| 0 \rangle = -F^2 B_0(1+O(m_q))\,.
\ee

The transformations under a chiral symmetry transformation
$g_L\times g_R~\in~SU(3)_L\times SU(3)_R$ are:
\ba
\rlabel{transformations}
U&\to&g_R U g_L^\dagger
\nonumber\\
\chi&\to&g_R \chi g_L^\dagger
\nonumber\\
u&\to& g_R u h^\dagger \equiv h u g_L^\dagger
\nonumber\\
R&\to& h R h^\dagger~\qquad{\rm for}\qquad R = \chi_+,u_\mu,V_\mu,V_{\mu\nu}
\nonumber\\
\Gamma_\mu&\to& h\Gamma_\mu h^\dagger +h\partial_\mu h^\dagger\,.
\ea
Eq. \rref{transformations} also serves as the definition of $h$, which depends
on $g_L$, $g_R$ and $\Pi$.

\section{Vector Meson Chiral Perturbation Theory}

\rlabel{HMET}
The Chiral Corrections we are interested in are long distance corrections
to a propagating ``bare'' vector meson\footnote{A similar formalism is of
course possible for axial vector mesons. The relevance for quark masses is
much smaller, since masses and mixings are much less well known in this sector}.
We can therefore calculate these corrections in a systematic fashion. We
cannot in a similar way calculate the chiral corrections to vector meson
decays of the kind: $V \to PP$ since that is inherently shorter distance.
``Hard'' loop corrections
to the vector meson masses will be included in the constants in the
effective lagrangians we will write. An example of this already at tree
level will be given in Sect. \tref{relativistic}.

It will also be nice to have explicit power counting present in the Lagrangian.
We therefore use a formalism similar to the Heavy Quark Effective
Theory\rcite{Georgi} and the one used for Baryons\rcite{JM,BKKM}. The main
part of the mass, the vector meson mass in the chiral limit, we will remove
explicitly by introducing a velocity $v$ with $v^2=1$ and the chiral mass $m_V$.
Vector momenta are then treated as residual momenta compared to $m_V v$.
I.e. $k_V = m_V v + p$ and we only refer to $p$ in the space time dependence.
We introduce the $3\times3$ matrix $W_\mu$ for the effective vectors:
\be
W_\mu =\left[
\begin{array}{ccc}
{\rho^0_\mu \over \sqrt{2}} + {\omega_\mu \over \sqrt{2}} & \rho_\mu^+ &
K_\mu^{*+} \\ \rho_\mu^- & -{\rho_\mu^0 \over \sqrt{2}} + {\omega_\mu \over
\sqrt{2}} & K_\mu^{*0} \\ K_\mu^{*-} & \overline K_\mu^{*0} & {\phi_\mu
}
\end{array} \right],
\rlabel{vectorMATRIX1}
\ee
Under the chiral symmetry group $W_\mu$ transforms as $W_\mu\to h W_\mu
h^\dagger$ with $h$ defined in Eq. \rref{transformations}.
The fields in \rref{vectorMATRIX1} are to be understood as only
containing annihilation operators. The creation operators are contained
in $W^\dagger$. In order to have a proper spin one field we impose the
condition
\be
v\cdot W = 0\,.
\ee
This condition is enforced by putting in the Lagrangian always
the combination $P^{\mu\alpha}W_\alpha$ and its Hermitian conjugate. Here
\be
\rlabel{defP}
P_{\mu\alpha} = g_{\mu\alpha} - v_\mu v_\alpha\,.
\ee
In the remainder this projector should be understood. For the calculation
presented in this paper it can be forgotten, since the $v\cdot W$ component
never contributes in any of the diagrams.

Let us now proceed to construct the relevant Lagrangians. The terms
at order $p^0$ are:
\be
\rlabel{L0}
{\cal L}_0 = \Delta_V \trace{W^\dagger_\mu W_\mu} 
+ x_0 \trace{W^\dagger_\mu}\trace{W^\mu}\,.
\ee
The first term, though allowed by the symmetry, can be removed from the
action through the choice of the reference momentum. The second term is
small because of Zweig's rule, we will therefore treat it as the same order
as the quark masses, ${\cal O}(p^2)$. The term at order $p$ is:
\ba
\rlabel{L1}
{\cal L}_1 &=& -i\trace{W^\dagger_\mu (v\cdot D)W^\mu} -
i x_1 \trace{W^\dagger_\mu}\trace{(v\cdot D)W^\mu} 
+\nonumber\\&&
i g \trace{\{W^\dagger_\mu , W_\nu\} u_\alpha}v_\beta \varepsilon^{\mu\nu\alpha\beta}
+i g' \left[\trace{W^\dagger_\mu} \trace{ W_\nu u_\alpha}
+\trace{W^\dagger_\mu  u_\alpha} \trace{ W_\nu}\right]
v_\beta \varepsilon^{\mu\nu\alpha\beta}\,\,.
\ea

For the purpose of counting, for constants in the lagrangian that are suppressed
by $1/N_c$, we will count them as order $p^2$. So the coefficient
$x_0$ needs to be kept in the loop diagrams but not $x_1$ and $g'$. Similarly,
for the $p^2$ Lagrangian below, the $1/N_c$ suppressed coefficients appearing in
loops we neglect.

For the higher orders we can use the lowest order equation of motion to remove
terms. So we will never encounter $(v\cdot D) W$ and $(v\cdot D)W^\dagger$.
We will also not explicitly mention terms that vanish as the external
vector and axial vector fields vanish. These never contribute to the
masses.
In addition, since we treat $x_1$ as a small quantity the effects
of this term can be reabsorbed in the other parameters by
the field redefinition
\be
\rlabel{remove}
W^\prime_\mu = W_\mu + \frac{x_1}{2} \trace{W_\mu}\,.
\ee
The field $W^\prime_\mu$ still transforms under the chiral group in the same
way as $W_\mu$. In fact the $x_1^2$ remainder can also be removed
by slightly changing the coefficient in \rref{remove}.

At order $p^2$ there are several terms. First there are those whose coefficients
are fixed by Lorentz-invariance. At the level of the HMET Lagrangian
these constraints can be directly derived using reparametrization
invariance\rcite{shift}.
The Lagrangian at order $p^2$ is then
\ba
\rlabel{L2}
{\cal L}_2 &=&
a_1 \trace{\{W_\mu^\dagger , W^\mu\}\chi_+}
+a_2 \trace{D_\mu W^\dagger_\nu D^\mu W^\nu}
+a_3 \trace{D^\mu W^\dagger_\mu D_\nu W^\nu}
\nonumber\\&+&
a_4 \trace{\left[\{D_\beta W^\dagger_\mu, W_\nu\}
              -\{D_\beta W_\nu, W^\dagger_\mu\}\right]
     u_\alpha} \varepsilon^{\mu\nu\alpha\beta}
+a_5 \trace{\{W^\dagger_\mu,W^\mu\}u_\alpha u^\alpha}
+a_6 \trace{W^\dagger_\mu u_\nu W^\mu u^\nu}
\nonumber\\&+&
a_7 \trace{\left(W^\dagger_\mu W_\nu+W_\mu W^\dagger_\nu\right)u^\mu u^\nu}
+a_8 \trace{\left(W^\dagger_\mu W_\nu+W_\mu W^\dagger_\nu\right)u^\nu u^\mu}
\nonumber\\&+&
a_9 \trace{W^\dagger_\mu u^\mu W_\nu u^\nu +W^\dagger_\mu u^\nu W_\nu u^\mu}
+a_{10}\trace{W^\dagger_\mu (v\cdot u)W^\mu(v\cdot u)}
+a_{11}\trace{\{W^\dagger_\mu,W^\mu\}\left(v\cdot u\right)^2}
\ea
In addition to the terms leading in $1/N_c$ of Eq. \rref{L2} we
also have
\ba
\rlabel{L2p}
{\cal L}^\prime_2 &=&
a_{12} \left(\trace{W^\dagger_\mu}\trace{\chi_+ W^\mu}
       +\trace{\chi_+ W^\dagger_\mu}\trace{W^\mu}\right)
+a_{13} \trace{W^\dagger_\mu W_\mu}\trace{\chi_+}
\nonumber\\&&
+a_{14} \trace{W^\dagger_\mu}\trace{W_\mu}\trace{\chi_+}
+a_{15} \trace{D^\mu W^\dagger_\nu}\trace{D_\mu W^\nu}
+a_{16} \trace{D^\mu W^\dagger_\nu}\trace{D^\nu W_\mu}
\ea
contributing to the vector masses directly.
The terms of Eqs. \rref{L0}, \rref{L1} and \rref{L2} are all those that
will appear within loops in the approximations used here.

The terms fixed by reparametrization invariance are:
\be
\rlabel{shiftresults}
a_2=\frac{-1}{2m_V}\qquad  a_4= \frac{g}{2m_V}\,.
\ee

We have not performed a full classification of the $p^3$ and $p^4$ terms,
this is work in progress, but only of those that can contribute to
the vector meson masses. We have shown the subleading ones in $1/N_c$.
These are all the terms needed to absorb the infinities occurring in the
present calculation.

There are no terms at tree level of order $p^3$ that contribute to
the vector masses.
The argument is as follows: In order to have $p^3$, we need three derivatives
or one insertion of quark masses and one derivative. To have an invariant term
contributing to the masses we also need a $W_\mu$, $W^\dagger_\mu$ and
Lorenz invariance requires an extra $v$. If the $v$ is contracted with the
vectors it vanishes because $v\cdot W=0$. If $v$ is contracted with a
derivative, it has to act on one of the vectors. This term then vanishes
by the equations of motion as mentioned above.

At $p^4$ we have four terms contributing to the masses:
\ba
\rlabel{L4}
{\cal L}_4 &=&
c_1\trace{W^\dagger_\mu \chi_+ W^\mu\chi_+}
+c_2\trace{\{ W^\dagger_\mu, W^\mu\}\chi_+\chi_+}
\nonumber \\&&
+ c_3\trace{ \chi_+\{ D_\mu W^\dagger_\nu,D^\mu W^{\nu}\} }
+c_4
\trace{\chi_+ \{ D^\mu W^\dagger_\mu,D_\nu W^{\nu}\}}
\ea
The last two terms in fact do not contribute to the masses at order $p^4$.
The reason is that the external momentum is proportional to a mass difference
that is itself of order $p^2$. So to the masses these terms only contribute
to order $p^6$. For a similar reason there are no contributions from terms
with four derivatives.

The $1/N_c$ suppressed terms are:
\ba
\rlabel{L4p}
{\cal L}_4 &=&
c_5 \trace{\chi_+}\trace{\chi_+ \{W^\dagger_\mu,W^\mu\}}
+c_6\trace{\chi_+ W^\dagger_\mu}\trace{\chi_+ W^\mu}
\nonumber\\&&
+c_7\trace{\chi_+\chi_+}\trace{W^\dagger_\mu W^\mu}
+c_8\left(\trace{\chi_+\chi_+W^\dagger_\mu}\trace{W^\mu}
   +\trace{\chi_+\chi_+W^\mu}\trace{W^\dagger_\mu}\right)
\nonumber\\&&
+c_9\trace{W^\dagger_\mu}\trace{W_\mu}\trace{\chi_+\chi_+}
+c_{10}\trace{\chi_+}\trace{\chi_+}\trace{W^\dagger_\mu W^\mu}
\nonumber\\&&
+c_{11}\trace{\chi_+}\left(\trace{\chi_+W^\dagger_\mu}\trace{W^\mu}
   +\trace{\chi_+W^\mu}\trace{W^\dagger_\mu}\right)
+c_{12}\trace{\chi_+}\trace{\chi_+}\trace{W^\dagger_\mu}\trace{W^\mu}
\ea

\section{Relation to Relativistic Vector Meson Lagrangians}
\rlabel{relativistic}

The usual parametrizations of vector mesons in chiral Lagrangians is done
in a relativistic formalism. There are various popular versions of this.
A review of some of them exists in \rcite{ULF}. A discussion relevant to
the relation between the different ways of parametrizing them can be found
in \rcite{TONI}. The relation in the context of the functional integral
approach can be found in \rcite{BP}.

Let us first treat the case of a single noninteracting Vector Meson.
The relevant Lagrangian is:
\be
\rlabel{LOREL}
{\cal L}^R_2 =
-\frac{1}{4} V^{\mu\nu}V_{\mu\nu}+\frac{1}{2}m^2 V^\mu V_\mu\,,
\ee
with $V_{\mu\nu}=\partial_\mu V_\nu - \partial_\nu V_\mu$.
This Lagrangian produces three propagating modes and one constraint equation.
The latter makes the non relativistic limit a little more tricky than just a 
naive identification.
We define first a parallel and perpendicular component of the vector
field with respect to the velocity $v$:
\be
V_\mu = P_{\mu\nu} V_\nu + v_\mu (v\cdot V) = \Vper_\mu+v_\mu \Vpar\,.
\ee
This is similar to the Heavy Quark Effective Theory where a projector
$(1+v\cdot\gamma)/2$ is introduced\rcite{Georgi}.
The Lagrangian of \rref{LOREL} then becomes
\ba
{\cal L}^R_2 &=& -\frac{1}{4} \Vper^{\mu\nu}\Vper_{\mu\nu}
+\frac{1}{2}m^2 \Vper^\mu \Vper_\mu
\nonumber\\&&
+(v\cdot\partial)\Vper_\mu \partial^\mu\Vpar+\frac{1}{2}m^2\Vpar^2
\nonumber\\&&
+\frac{1}{2}\left( (v\cdot\partial)\Vpar\right)^2
-\frac{1}{2}\left(\partial_\mu \Vpar\right)^2\,.
\ea
We can now split both $\Vper$ and $\Vpar$ in its creation and annihilation
parts (keeping in mind the projector acting on the effective field):
\ba
\Vper_\mu &=& \frac{1}{\sqrt{2 m_V}}\left[e^{-i m_V v\cdot x} W_\mu +
  e^{i m_V v\cdot x}W^\dagger_\mu\right]\,;
\nonumber\\
\Vpar &=& \frac{1}{\sqrt{2 m_V}}\left[e^{-i m_V v\cdot x}W_\parallel +
  e^{i m_V v\cdot x}W^\dagger_\parallel\right]\,.
\ea
We can then identify $m_V = m + \cdots$ where the dots are a small quantity.
We also assume that the residual momentum dependence described by $W_\mu$,
$W_\parallel$ and Hermitian conjugates, is small compared to $m_V v$.
This is where we restrict to the one vector meson sector.

The only terms in the action that are still proportional to $m_V$ is
$(m_V W_\parallel^\dagger\cdot W_\parallel)/2$. So, in this limit we
precisely have $v\cdot V=0$ or the constraint we assumed in the previous
section $v\cdot W = 0$. The symbol $W$ used in the previous section
always included the projector $P$ of \rref{defP}; i.e. it is $W_\perp$.
One could also alternatively integrate out the $\Vpar$ component in a
functional integral language. Here we will just remove it using the
equations of motion:
\be
W_\parallel =
\frac{-i}{m_V}\partial_\mu W_\perp^\mu
+\frac{1}{m_V^2}\left\{\partial_\mu(v\cdot\partial)W_\perp^\mu
+\left[\partial^2-(v\cdot\partial)^2\right] W_\parallel\right\}\,.
\ee
This equation can simply be solved iteratively.
If applied to the kinetic terms this leads to the constraints given
in Eq. \rref{shiftresults}.

In practice we can choose slightly different values of $m_V$ to
do the reduction to the HMET. This is precisely the freedom that leads to
the presence of the term proportional to $\Delta_V$ in Eq. \rref{L0}.

An alternative procedure, that is in practice somewhat easier to implement,
is to use tree level matching. We calculate certain processes with the
relevant kinematics both in the relativistic formulation and in the HMET
formulation. We then determine the HMET constants in terms of the
constants in the relativistic Lagrangian by requiring equality between the
two formulations. 
Here one has to watch out somewhat. The equivalence is at the level of
$S$-matrix elements so when comparing Green functions we can only easily compare
for the perpendicular components and one should remember the extra
factors of $1/(2 m_V)$ because the relativistic and HMET fields are differently
normalized.

\section{Coefficients in the Lagrangian}
\rlabel{coefficients}

The total number of parameters that contribute to order $p^4$ to the
vector meson masses is very large. It is obviously too large to be
fitted simply from the data on the masses alone. In our numerical results
we will therefore use several estimates of the parameters.

The most important ones are the extra $1/N_c$ assumptions used. Here we
add a suppression factor of $p^2$ for each factor
of $1/N_c$. The loops themselves are in principle also suppressed by factors
of $1/N_c$. But large logarithms and combinatorial factors can in
practice make up for this extra loop suppression. We therefore still take
the loops into account.
For the parameters this means:\footnote{As mentioned earlier,
derivatives on external vector field count as $p^2$ for the calculation
of the masses.}
\be
g'=a_{14}=a_{15}=a_{16}=c_5=\cdots=c_{12}=0\,.
\ee
Here equal to zero means that they do not contribute to the masses when the
$1/N_c$ expression is taken into account.
$x_0$ should be included inside loops and at tree level. 
$x_1$ is not physically relevant as described earlier.
$a_{12}$
and $a_{13}$ only contribute at tree level. $a_{13}$ is like a vector
meson sigma term. On the masses it only contributes just like
$\Delta_V$ and is thus of no relevance here.
In the end there are two possibly relevant $1/N_c$ parameters,
$x_0$ and $a_{12}$.

First the $vector-vector-pseudoscalar$ terms. Here we only need $g$.
It was estimated in \rcite{ELISABETH} using the chiral quark model
with a value of $g=0.375$. Assuming $g_{A{\rm quark}}=1$, 
this changes to $g=0.5$.
In \rcite{PETERHANS} the value of $g$ was fixed using a VMD argument for
$\omega\to\pi\gamma$, this yielded $g=0.32$ in reasonable agreement with
the previous estimate. A double VMD estimate from $\pi^0\to\gamma\gamma$
leads to essentially the same result.
The ENJL prediction for this vertex\rcite{XIMO} is lower $g=0.27$.
Fitting to $\Gamma(\rho^0 \to
\gamma \pi^0)$,
$\Gamma(\rho^+ \to \gamma \pi^+)$ and $\Gamma(\rho\to\pi\pi)$,
together with various assumptions about VMD and the KSRF relation,
leads also to values roughly within the above range.
We will discuss the implications for the vector meson masses
of these and other values in 
Sec. \tref{numerics}. 

The terms with more derivatives and vector fields only are as described
earlier fully determined by Lorenz invariance, $a_2$ and $a_4$.
$a_3$ never contributes to the masses.

The $vector-vector-pseudoscalar-pseudoscalar$ interaction terms are more
difficult to obtain. In \rcite{TONI} a good description of Vector
meson
phenomenology is given by the model III:
\be
{\cal L}_{III} = -\frac{1}{4} \trace{\bar{V}_{\mu\nu}^2} +\frac{1}{2}m_V^2
\trace{\left(\bar{V}_\mu-\frac{i}{\hat{g}}\Gamma_\mu\right)^2}\,.
\ee
Here $\bar{V}_\mu$ transforms nonlinearly under the chiral group
as $\bar{V}_\mu\to h \bar{V}_\mu h^\dagger +i/\hat{g}~h\partial_\mu h^\dagger$
and
\be
\bar{V}_{\mu\nu} =
\partial_\mu \bar{V}_\nu-\partial_\nu \bar{V}_\mu-
 i \hat{g}\left[\bar{V}_\mu,\bar{V}_\nu\right]\,.
\ee
In this model there are three diagrams possibly contributing to $\pi V_\mu$
scattering. These are depicted in Fig. \tref{figrel}.
The effects of these have to be described by the pointlike interaction
in the HMET.
We have chosen this model 
because most properties of Vectors and Axial vectors are well described by the
 Yang-Mills type models. 
The type ``hard'' contributions of Fig. \tref{figrel}a cannot be simply
described using equations of motion as was done in the previous section.
For these we use the matching procedure.
The internal propagator can both be a pseudoscalar or an axial vector.
Similarly in Fig. \tref{figrel}c the vector in the vertical line has a small
momentum, not a small deviation from $m_V v$.
Diagram Fig. \tref{figrel}c reproduces the pion interactions from
the kinetic term and the $a_2$ term of the HMET in addition to:
\be
a_3 = -a_7 = a_8 = \frac{1}{2 m_V}\,.
\ee
This doesn't contribute to the masses.
The diagram of Fig. \tref{figrel}a is reproduced in the HMET via
\be
-a_8 = a_9 = \frac{m_V}{16 g^2 F^2}\approx \frac{1}{4 m_V}\,.
\ee
\begin{figure}
\begin{center}
\leavevmode\epsfxsize=12cm\epsfbox{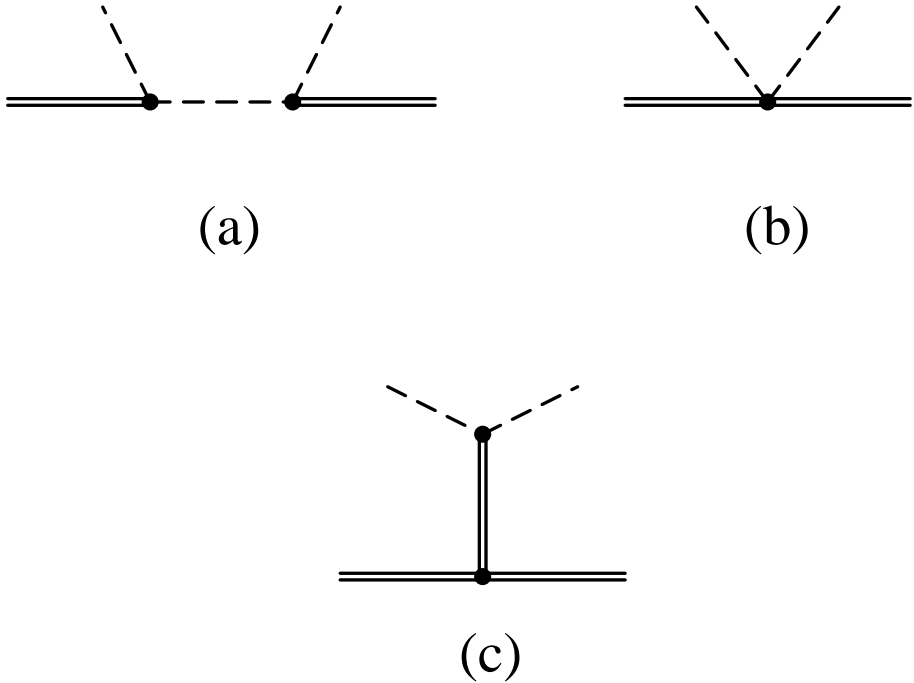}
\end{center}
\caption{\rlabel{figrel}The three possible relativistic diagrams
contributing to pseudoscalar-vector scattering.}
\end{figure}

Higher momentum dependences of the $\pi V$ scattering can be produced
by the axial vector intermediate states. The $\bar{V}_\mu A_\nu \pi$ vertex
in this model can be described by the two terms
\be
i A_1 \trace{\left(\bar{V}_\mu-\frac{i}{\hat{g}}\Gamma_\mu\right)
\left[ u_\nu,D_\mu A_\nu\right]}
+i A_2 \trace{A_\mu\left[u_\nu,\bar{V}_{\mu\nu}\right]}\,.
\ee
in addition to the kinetic terms for the axial nonet. The field $A_\mu$
transforms here as $R$ in Eq. \rref{transformations}.
In the HMET they can be described via
\ba
a_8 = -a_9 &=& -\frac{m_V}{2 m_A^2}\left( A_2\right)^2\nonumber\\
a_{10} = -2a_{11}&=&\frac{m_V}{(m_A^2-m_V^2)}\left(A_1+A_2\right)^2\,.
\rlabel{A1contr}\ea
The coefficients $A_1$ and $A_2$ should then be chosen to reproduce the
rather uncertain $a_1$ width.

The above model did not contain much internal vector meson properties.
As an example of a model that does have some internal structure we will use
the ENJL model. For a review see \rcite{PHYSREP}. The coefficients needed
here were obtained in \rcite{EDUARD} and \rcite{XIMO}. In order to obtain the
$a_i$ from there we have to perform the HMET reduction. We also have to include
the diagrams of Fig. \tref{figrel}, keeping in mind that
in \rcite{EDUARD,XIMO} a different vector representation was used where
the vector-pseudoscalar-pseudoscalar vertex contains three derivatives.
What it in the end corresponds to is that we have the above
contributions in terms of $\hat{g} = 1/(2\sqrt{2}g_V)$ and
$A_1 = A^{(2)}$; $A_2 = A^{(3)}$ and in addition terms coming from
the pointlike relativistic contribution of Fig. \tref{figrel}b.
These extra terms are
\ba
a_5^E=\frac{\delta^{(1)}_V}{4 m_V};&&\,
a_6^E=\frac{\delta^{(2)}_V}{2 m_V}\,;\qquad
a_7^E=\frac{\delta^{(3)}_V}{4 m_V}\,;
\nonumber\\
a_8^E=\frac{\delta^{(4)}_V}{4 m_V};&&\,
a_9^E=\frac{\delta^{(5)}_V}{2 m_V}\,;
\ea
The symbols $g_V$, $A^{(2)}$, $A^{(3)}$ and $\delta^{(1,\ldots,5)}$ are
defined and their values in terms of the ENJL parameters given in \rcite{XIMO}.
For the contributions to the masses, $a_7$ and $a_8$ only appear in the
combination $a_7+a_8$.
To obtain the numerical values we can use the ENJL relations:
\ba
\rlabel{numbers1}
-\delta^{(1)}_V &=& \delta^{(2)}_V=\delta^{(3)}+\delta^{(4)}=-2\delta^{(5)}
=\frac{f_A^2}{2f_V^2}\approx 0.12
\nonumber\\
A^{(2)} &=& -\frac{1}{2}\frac{f_A}{f_V}\approx -0.24
\nonumber\\
A^{(3)} &=&A^{(2)}+\frac{2}{f_A f_V}\left(L_9+L_{10}\right)\approx
-0.075\,.
\ea
The numerical values correspond to $f_V = 0.20$, $f_A = 0.097$
\rcite{PHYSREP} and
$L_{9}+L_{10}=1.6\cdot10^{-3}$\rcite{TWOLOOP}.

The remaining constants to be estimated are $a_1$, $c_1$ and $c_2$. For these
we will use an ENJL model estimate. The Yang-Mills like models normally
just assume these terms to be zero. These are the vector meson masses
in the large $N_c$ limit. We could in principle determine those
from the measured meson masses but we will here also produce an ENJL
calculation of them. In the ENJL model the two-point functions were
well described by a VMD picture, see \rcite{PHYSREP,XIMOHANS}. We will therefore
do the same thing. We will fit the two-point function
$\Pi^{(0+1)}_V(-q^2)$ to $Z_V/(M_V^2-q^2)$. 
We do this fit in the Euclidean region to remove the artifacts due to
non-confinement in the ENJL model.
We observe that a good
fit can be obtained with $Z_V$ independent of the quark masses and
will use the fitted $M_V^2$ to obtain the
coefficients $a_1$, $c_1$ and $c_2$. Notice that we fit the squared of the mass
and not the mass itself, like one should expect in an effective theory.
With this we take into account the resummation of the terms $1/m_V$ 
suppressed inside these coefficients. The values chosen for the
ENJL parameters are $G_S = 1.216$, $G_V = 1.263$ and $\Lambda_\chi=1.16~\mbox{GeV}$
which gives a good fit to a large number of hadronic parameters.
We have also used the way suggested in \rcite{XIMOHANS} to determine the
vector masses. 
This is another recipe to remove the pole at $q^2=0$ in 
$\Pi^{(1)}_V$. The differences are within the quoted errors.
We have also done the fit using various ranges of the quark masses
and various ranges of $q^2$. The results are all within the quoted
values.
We will use the values
\be
\rlabel{numbers2}
a_1 = (0.15\pm0.01)~\mbox{GeV}^{-1}\qquad c_1=(-0.08\pm0.03)~\mbox{GeV}^{-3}\qquad
c_2 = (-0.06\pm0.02)~\mbox{GeV}^{-3}
\ee
The vector mass in the chiral limit obtained from the ENJL model with the
above parameters is about $0.630~\mbox{GeV}$. The model uncertainty on these
estimates increases the error on $a_1$ to a somewhat larger value.\\
\section{Determination of the Masses}

\rlabel{calculation}
The method we will use to determine the masses is to compute the
inverse propagator and then we look for its zeros.
As mentioned earlier we perform the calculation fully in the HMET.
We are free to choose the external momentum in a way that simplifies
the calculation. Our choice is that the external momentum is always
proportional to the chosen velocity $v$. This together with the constraints
$v\cdot W=0$ removes a lot of irrelevant contributions.
A disadvantage of the HMET are that there are more possible diagrams,
corresponding to the extra operators with fixed coefficients. 
In the relativistic formulation we only have two possible diagrams.
Then we have no simple powercounting because the large
mass appears in the nonanalytic parts. In the
nonrelativistic case we have more diagrams.
These are depicted in
in Fig. \tref{figdiag}.
\begin{figure}
\begin{center}
\leavevmode\epsfxsize=12cm\epsfbox{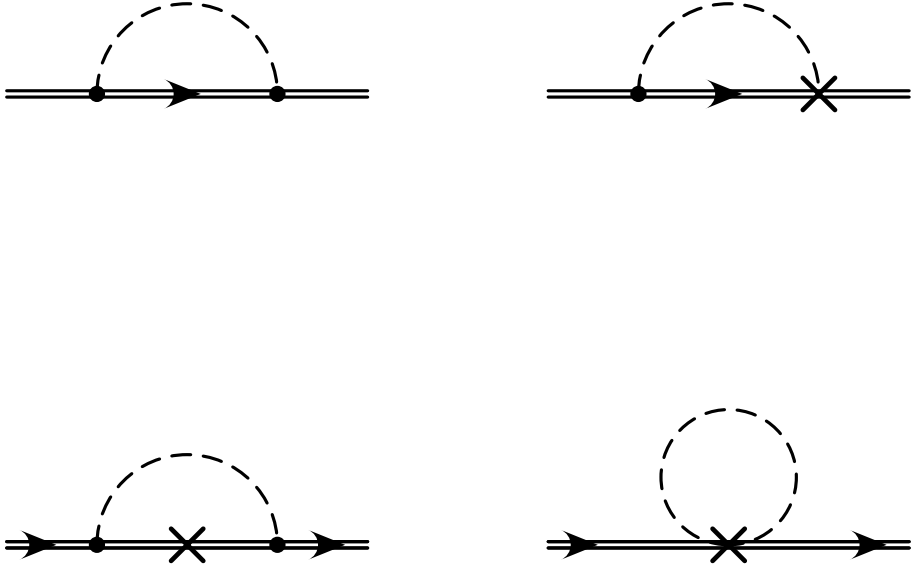}
\end{center}
\caption{\rlabel{figdiag} The loop diagrams contributing to the masses in the
HMET formulation. The dot is an ${\cal L}_1$ vertex. The crosses
are ${\cal L}_2$ vertices.}
\end{figure}

As argued before we only have to include the effects of the terms up to $p^2$
inside the loops. In practice this means that we have to diagonalize the masses
including the terms proportional to $x_0$ and $a_1$ inside the sunrise type
diagrams, depicted in the first three diagrams of Fig. \tref{figdiag}.
We also diagonalize the lowest order pseudoscalar mass term to
take $\pi^0-\eta$ mixing into account.

We proceed in the following way. We define a set of external vector fields
with $W_\mu = W_{ext\mu}^a T^a_{ext}$ that corresponds to the usual
isospin basis. We then define a 2nd set of ``internal'' vector fields
$W_\mu = W_{int\mu}^a T^a_{int}$ that diagonalize the lowest order mass terms,
i.e. the $\Delta_V$, $a_1$, $x_0$ and $\delta_1$ (see next Section) terms.
Similarly we define a set of pseudoscalar meson fields $\pi^M$ that
diagonalize their lowest order mass term, $\Pi = \pi^M T^M$.

In terms of these the sunrise diagrams depicted the first three 
diagrams in Fig. \tref{figdiag}
as a function of the incoming flavour $a$ and outgoing vector $b$ with
residual momentum $p_\mu = p_{ext}v_\mu$ can be described via:
\be
\sum_{M=1,8}\sum_{c=1,9}\frac{4g^2}{F_\pi^2}
\trace{\{T_{ext}^a,T_{int}^{c\dagger}\}T^M}
\trace{\{T_{ext}^{b\dagger},T_{int}^c\}T^{M\dagger}}
i g_{\mu\nu} K(p_{ext},\Delta m_c, m_M)
\ee
The function $K$ is defined as follows:
\be
\rlabel{vectordiag}
i K g_{\mu\nu} + i L v_\mu v_\nu =
\int\frac{d^dq}{(2\pi)^d}\frac{q_\mu q_\nu}{q^2 - m_M^2}
\frac{1}{v\cdot q + p_{ext} -\Delta m_a}
\left( 1 + \frac{p_{ext}}{m_V}-\frac{m_M^2}{2m_V}
\frac{1}{v\cdot q + p_{ext} -\Delta m_a}
\right)\,.
\ee
$K$ can be easily constructed from the integrals given in
App. \tref{integrals}. $\Delta m_a$ is the vector meson mass minus the
chiral limit value, i.e. $\Delta m_a = m_a - m_V$. $m_M$ is the mass of
the pseudoscalar.
We have disregarded in these expressions 
terms of higher orders found in the integrals, i.e $
p_{ext}^2$ and $p_{ext} \Delta m_a$, their numerical effects are found
to be small and mainly affects de $\phi$. For the $\phi$ we actually
find a second zero in the inverse propagator for some of the parameter
sets. Adding the omitted higher order terms in the sunrise diagrams
moves this extra zero beyond the range of validity of CHPT as discussed
in Sect. \tref{numerics}.
These extra terms correspond to replacing $m_M^2$ in the last term
of $K$, Eq. \rref{vectordiag}, by $m_M^2+2 p_{ext}\Delta m_a-p_{ext}^2$.

The tadpole contributions, last diagram in Fig. \tref{figdiag} can be given by
\ba
\rlabel{tadpole}
&ig_{\mu\nu}&\sum_{M=1,8}
\frac{a_1}{2F_\pi^2}\mu_M \trace{\{T_{ext}^a,T_{ext}^{b\dagger}\}
\{T^M,\{T^{M\dagger},\chi\}\}}
\nonumber\\
&-ig_{\mu\nu}&\frac{2}{F_\pi^2}\sum_{M=1,8}m_M^2\mu_M
\left[a_5+\frac{1}{d}\left(a_7+a_8+a_{11}\right)\right]
\trace{\{T_{ext}^a,T_{ext}^{b\dagger}\}T^M T^{M\dagger}}
\nonumber\\
&-ig_{\mu\nu}&\frac{2}{F_\pi^2}\sum_{M=1,8}m_M^2\mu_M
\left[a_6+\frac{1}{d}\left(2a_9+a_{10}\right)\right]
\trace{T_{ext}^a T^M T_{ext}^{b\dagger} T^{M\dagger}}\,.
\ea
The traces can be easily performed but lead to rather lengthy expression.
In App. \tref{formulas} we have given the expressions in the limit where
internal lines are the nonmixed states. The numerical difference,
especially the part due to the $\pi^0-\eta$ mixing, is rather small,
less than half an MeV for all masses.

To equations \rref{vectordiag} and \rref{tadpole} we still have to
add all tree level contributions to obtain the inverse propagator
to order $p^4$. The corrections to the masses of the 
$\rho^+,~K^{*0},~K^{*+}$ are given by values of $p_{ext}$ for which the
corresponding amputated two point function vanishes. For the 
$\rho^0,~\omega,~\phi$ the situation is slightly more complicated, 
because these mix with each other. Here the corrections to the masses
are given by the values of $p_{ext}$ for which the determinant of the
matrix of the amputated two point function vanishes.
Once we have found the three solutions, we have to identify the
particle to
which they correspond. This can be done by analyzing the null
eigenvectors at the value of $p_{ext}$ where the determinant vanishes. 
These null eigenvectors define the 
physical basis for the $\rho, \omega$ and $\phi$, and they do not
necessarily form an orthogonal set since the physical states appear at
different values of $p_{ext}$.

\section{Electromagnetic contributions to the Masses}

\rlabel{electromagnetic}
The subject of the electromagnetic contributions to masses is in fact very old,
an early attempt is in \rcite{singer}. In \rcite{PETERHANS} a
matching calculation was performed. We will use the results of that
paper here. 
The main electromagnetic effects can be described by the three terms
\be
\delta_1\trace{QW^\dagger_\mu}\trace{QW^\mu}
+\delta_2\trace{\left[Q,W^\dagger_\mu\right]\left[Q,W^\mu\right]}
+\delta_3\trace{Q^2\{W^\dagger_\mu,W^\mu\}}\,,
\ee
with as numerical values\rcite{PETERHANS}
\be
\rlabel{numbers3}
\delta_1=0.27\pm0.11~\mbox{MeV}\qquad\delta_2=-1.4\pm0.4~\mbox{MeV}{\rm~and~}
\delta_3 =0.27\pm0.11~\mbox{MeV}\,.
\ee
The numerical results from \rcite{XINESOS} cannot be easily
compared. The model used there makes rather drastic assumptions about the
high energy behaviour of the various formfactors needed. The bad matching
found in \rcite{PETERHANS} should however be kept in mind.
There are several effects included in \rcite{singer,XINESOS} that
in the language of \rcite{PETERHANS} are higher order. The main effect
of these is the $\pi\gamma$ intermediate state.
The effect of this we model in the present work by including the $\delta_1$
term in the diagonalization used in the previous section
and by using the measured vector meson masses inside the loops.

Another possibly large effect is the electromagnetic part of the pseudoscalar
meson masses in the loops. This is in fact the leading quark mass effect
in the electromagnetic corrections\rcite{PETERHANS}. Here it is included
by using the physical pseudoscalar masses within the loops\footnote{The
effects on the mixings of pseudoscalars is higher order and we neglect it.}.

\section{Numerical Results}
\rlabel{numerics}

As a general inputs we choose $m_u = 4.273$~MeV, $m_d = 7.727$~MeV 
and $m_s = 146.0~\mbox{MeV}$ as well as
$B_0 = m_{\pi^0}^2/(m_u+m_d)$. These masses are compatible with the small
deviation from the Gell-Mann-Oakes-Renner relation and the quark mass
$\hat{m} = m_u+m_d$
determined in \rcite{BPR}.
The other values are found via the ratios
quoted in \rcite{Leutwyler}. We also take the 
coefficients given in Eqs. \rref{numbers1}, 
\rref{numbers3} as fixed parameters. We also fix the $a_1$ mass appearing
in Eqs. \rref{A1contr} via $m_A = \sqrt{2}\ m_V$ and
choose as a subtraction point $\mu=0.77$~GeV. That leaves
7 parameters to vary, $m_V$, $x_0$, $a_1$, $a_{12}$, $c_1$, $c_2$ and $g$.
We evaluate the masses in different scenarios:

Scenario $I$, we use the ENJL estimates for $a_1$, $c_1$, $c_2$, and the set of
parameters given in Eq.
\rref{numbers2}.
We also use $x_0 = 6.5$ MeV and $g=0.32$ \rcite{PETERHANS}, with
$m_V = m_\rho$. We found a large $\omega\phi$ mixing with
a large  $p_{ext}$ dependence. There is also a very large negative
correction to the $\phi$-mass resulting in the $\phi$ being the lightest
state. Most of these problems are caused by the very large contribution from
the sunrise diagrams. We can use $x_0$ and $a_{12}$ to try to solve these
problems. $x_0$ would lead to a very large $\omega\rho^0$ mass difference
so we will use $a_{12}$ to cancel the $\omega\phi$ mixing.
Requiring the $\omega \phi$ mixing to be small in a reasonable
range of $p_{ext}$ leads to $a_{12} \sim 0.16$ GeV$^{-1}$.

A very similar result can be obtained with $g=0.27$ as suggested by the ENJL
model. It has the masses somewhat better than scenario I but there is no
qualitative improvement, we therefore do not discuss it further.

This result is  not in good agreement
with the experimental data
as shown in Table \tref{tableout}.  We should keep in mind that the values
for $a_1$, $c_1$, $c_2$ and $a_{12}$ we have used are rough estimates,
and that $x_0 = 6.5$ MeV is a tree level value which can receive
corrections mainly from the sunrise diagram at ${\cal O}(p^3)$.

In the remainder we therefore use a least squares fit to determine viable
sets of parameters. We have minimized the following function:
\newcommand{\ratios}[2]{\left(\frac{#1}{#2}\right)^2}
\ba
F_{min}&=&
\ratios{m_{\rho^+}-m_{\rho^0}}{1~\mbox{MeV}} +
\ratios{(m_{\rho^+}+m_{\rho^0})/2}{10~\mbox{MeV}} +
\ratios{(m_{K^{*+}}+m_{K^{*0}})/2}{20~\mbox{MeV}} 
\nonumber\\&&+
\ratios{m_{\rho^0}-m_{\omega}}{1~\mbox{MeV}} +
\ratios{m_{K^{*+}}-m_{K^{*0}}}{1~\mbox{MeV}} +
\ratios{m_{\phi}-m_{\rho^+}}{40~\mbox{MeV}}\,.
\rlabel{fitfunction}
\ea

The masses should be understood here as the difference of the CHPT calculation
and the measured values. All data are from Ref. \rcite{DPG} except for the
$\rho^+-\rho^0$ mass difference\rcite{aleph}.
Here we have chosen the errors so that the quantities mainly determined by
$m_s$ are allowed a larger variation. We will then have to check afterwards
if the found sets have a reasonable convergence in the various orders.
\begin{table}
\begin{center}
\begin{tabular}{|c|c|c|c|c|c|c|}
\hline
Scenario           & I      & II      & III    & IV       & V      & VI\\
\hline
$m_V$ (GeV)        & 0.7685 & 0.84304 & 0.85801 & 0.84250 & 0.74516 & 0.75935\\
\hline
$x_0$ (GeV)        & 0.0065 & 0.00578 & 0.00756 & 0.00581 & 0.0195  & 0.0198\\
\hline
$a_1$ (GeV$^{-1}$) & 0.15   & 0.16371 & 0.12484 & 0.16926 & 0.12166 & 0.11990\\
\hline
$c_1$ (GeV$^{-3}$) & $-$0.08  & 0.13582 & 0.15531 & 0.13718 & 0.17327 & 0.17307\\
\hline
$c_2$ (GeV$^{-3}$) & $-$0.06  & $-$0.12964 & $-$0.08406 & $-$0.13457& $-$0.0373 & $-$0.03873\\
\hline
$a_{12}$ (GeV$^{-1}$)& 0.16 & 0.10284 & 0.10070 & 0.10279 &$-$0.01937 & $-$0.00937\\
\hline
$g$                & 0.32   & 0.25749 & 0.26507 & 0.25863 & 0.11926 & 0.14352\\
\hline
\end{tabular}
\end{center}
\caption{The input parameters for the various scenarios. The units are
given in the first column}
\rlabel{tablein}
\end{table}

\begin{table}
\begin{center}
\begin{tabular}{|c|c|c|c|c|c|c|}
\hline
Scenario & I & II & III & IV & V & VI\\
\hline
$\rho^+$ (GeV) & $-$0.1666 & $-$.00001 &0.0001&   -    &     -   &    - \\
\hline
$K^{*+}$ (GeV) & $-$0.1658 & +0.0009 &$-$0.0023 & 0.0004 & $-$0.0003 &
 - \\
\hline
$K^{*0}$ (GeV) & $-$0.1708 & +0.0009 &$-$0.0025 & 0.0004 & $-$0.0003 &
 - \\
\hline
$\rho^0$ (GeV) & $-$0.1665 & $-$0.0005 & 0.0001 &$-$0.0005& $-$0.0001 & - \\
\hline
$\omega$ (GeV) & $-$0.1638 &     -   & 0.0001  & $-$0.0005 & $-$0.0001 & - \\
\hline
$\phi$ (GeV)   & $-$0.3177 &     -   & 0.0003  & $-$0.0009 & +0.0003 &
 +0.0001 \\
\hline
$\omega\rho^0$($\rho^0$) (GeV) & $-$0.0023 & $-$0.0024 & $-$0.0016 & 
$-$0.0025 & $-$0.0016 & $-$0.0014 \\
\hline
$\rho^0\phi$ ($\phi$) (GeV) & +0.0004 & +0.0009 & +0.0008 & +0.0008 &
 +0.0006 & +0.0005 \\
\hline
$\omega\phi$ ($\omega$) (GeV) & $-$0.035 & $-$0.042 & $-$0.048 & $-$0.043 &
 $-$0.071 & $-$0.072 \\
\hline
$\omega\phi$ ($\phi$) (GeV) & $-$0.056 & +0.015 & +0.009 & +0.011 &
 $-$0.055 & $-$0.050 \\
\hline
$\sqrt{F_{min}}$ & 18 & 0.49 & 0.19 &0.52& 0.06 & 0.017 \\
\hline
\end{tabular}
\end{center}
\caption[The]{The deviation from the observed vector masses.
 for the various scenarios. Also given is the total standard deviation,
this $\sqrt{F_{min}}$ defined in Eq. \protect{\rref{fitfunction}}.
The units are
given in the first column. A dash means agreement to better than 4 digits.}
\rlabel{tableout}
\end{table}
We will present five different fits, all of which are acceptable.
The input parameters can be found in Table \tref{tablein} and the resulting
deviation from the observed masses in Table \tref{tableout}.
In that table we also present the predictions for the various mixings.
These are evaluated as the two-point function in the isospin basis
at the relevant value of the momentum.
Notice that in most fits these were not input so they present genuine
predictions. The $\omega \phi$ mixing has a large $p_{ext}$ dependence.
We have given it both at the $\omega$-mass and the $\phi$-mass. The $\rho\phi$
mixing is more stable and we quote it only at the $\phi$-mass. For definiteness
we have quoted the $\rho^0\omega$ mixing at the $\rho^0$-mass. Using the 
$\omega$-mass instead is not visible with the precision quoted.

Scenario II is a minimum found with a fairly large value of
$g$. All masses can be fit within the errors assumed in Eq. 
\rref{fitfunction}. All parameters are of the expected order of magnitude
here, the main change is the flip in sign of $c_1$.
As mentioned earlier, there is a second spurious zero in the determinant
of the inverse two-point functions of the neutral states about 60~MeV
above the $\phi$-mass. In order to check the sensitivity to higher orders
of this phenomenon and the possible variations of the parameters we have
also performed a fit with the higher order parts of the sunrise diagrams
included. This pushes the spurious zero to about 230~MeV above the
$\phi$-mass and varies the input parameters within less than one standard
deviation from the previous fit. That case is in the tables as scenario III.

A much better minimum can be obtained for a value of $g$ around $0.1$.
This is quoted in scenario V and VI. The latter is again with the higher order
parts of the sunrise diagrams included.

As can be easily seen we obtain reasonable
agreement with the observed values for
all the mixings as derived in \rcite{PETERHANS}.
These were:
\be
\rho^0\omega = -2.5~\mbox{MeV};\qquad
\rho^0\phi(\phi) = 0.3~\mbox{MeV}\qquad \mbox{and}\qquad
\omega\phi(\phi) = (8\mbox{ to }14)~\mbox{MeV}\,.
\ee
We should keep in mind that the latter two have possibly large final
state corrections.

The agreement with the $\rho^0\omega$ mixing in scenario III is not so good.
We have therefore added to Eq. \rref{fitfunction} also
\be
\ratios{\rho^0\omega(\rho^0)+0.00254~\mbox{GeV}}{0.0002~\mbox{GeV}}\,.
\ee
Including this then gave a minimum within the errors expected of the
input parameters. This we have given as scenario IV.

We have not discussed the $\mu$ dependence in the two-point
function because this can be absorbed in a change in the parameters
in principle if we include the $1/N_c$-suppressed $p^4$ terms as well.
In practice we have to fit the parameters anyway. We have checked that
putting $\mu=0.9$~GeV instead we can get as good as a fit as scenario III
and IV with similar changes in the input parameters.

To see the convergence of the series for the various quantities we show
in Table \tref{tableconv} the various contributions corresponding to scenario
IV. They are ordered by the order of the contributions. The sunrise diagrams
contain pieces of order $p^3$ and higher. The $a_2$ term is a term with two
derivatives but as explained earlier these count as $p^4$ for the calculation
of the vector masses. For the masses of the neutral states we have quoted
the various contributions to the unmixed inverse two-point function
at the correct mass. The effect of the mixing on the masses can be seen
by comparing the total line to the real masses as obtained in
Table \tref{tableout} for scenario IV.
\begin{table}
\begin{center}
\begin{tabular}{|c|r|r|r|r|r|r|r|r|r|}
\hline
 &$\rho^+$&$K^{*+}$&$K^{*0}$&$\rho^0$&$\omega$&$\phi$&$\rho^0\omega(\rho)$&
$\rho^0\phi(\phi)$&$\omega\phi(\phi)$\\
\hline
$a_1$ &12.3&154.5&158.0&12.3&12.3&300.1&$-$3.55&0&0\\
$x_0$ &0 & 0 & 0 & 0 &11.6 & 5.8&0&0&8.2\\
\hline
Sun   &$-$140.8&$-$126.2&$-$127.2&$-$140.8&$-$143.6&$-$329.5&3.26&2.6&$-$90.3\\
\hline
$a_1$-tad&26.3&53.2&53.6&26.3&26.3&80.5&$-$0.44&0&0\\
$a_i$-tad&30.1&70.7&71.7&30.1&29.0&58.1&$-$0.28&0.4&$-$41.1\\
\hline
$a_{12}$& 0  & 0    & 0    & 0 &15.0&182.3&$-$2.16&$-$1.5&134.2\\
$c_1$&  0.2&3.2&5.7&0.2&0.2&107.8&$-$0.10&0&0\\
$c_2$&$-$0.4&$-$105.9&$-$106.1&$-$0.4&$-$0.4&$-$211.6&0.20&0&0\\
$a_2$&$-$3.2&$-$1.5&$-$1.7&$-$3.3&$-$2.2&$-$18.4&0&0&0\\
\hline
em &  1.5 & 1.5 & 0.06& 1.5 &0.3 & 0.4 & 0.54 & $-$0.6&$-$0.2\\
\hline
Total & $-$74.0&49.5 & 54.0 & $-$74.0 &$-$51.4 & 175.5& $-$2.53&0.8&10.8\\
\hline
\end{tabular}
\end{center}
\caption[various]{The various contributions
to masses and mixings for scenario IV.
All units here are MeV. The contributions are labeled by the coefficient
in front of the vertex for the tree level contributions. The loop diagrams
are the sunrise diagrams (Sun), the tadpoles with the $a_1$ vertex
($a_1$)-tad and the other tadpoles ($a_i$-tad). The electromagnetic corrections
from \protect{\rcite{PETERHANS}} (em) and the sum are in the final lines.}
\rlabel{tableconv}
\end{table}
As can be seen the convergence for most quantities is acceptable but the
$\phi$-mass is very slowly converging.

\section{Conclusions}
\rlabel{conclusions}

We have in this paper extended the calculation of the vector meson masses
in CHPT beyond $p^3$\rcite{ELISABETH} and included also the isospin breaking
effects. The problem that the $p^3$ corrections break several observed
regularities can be repaired by the introduction of the higher orders.
The large breaking of the equal space relation, $m_{\phi}-m_{K^*} =
m_{K^*}-m_\rho$, and the large $\omega-\phi$ mixing are also solved.
In addition the problem of the $K^{*+}-K^{*0}$ mass difference\footnote{In
Ref. \protect{\rcite{XINESOS}} the latter problem was discussed at order $p^2$
including the electromagnetic corrections and a different solution proposed.}
with usual quark mass ratios can be resolved.
The parameters we obtain from  a fit to the masses are very good with
reasonable choices of all the parameters. The orders of magnitude of the
various parameters expressed in GeV are all natural with the possible exception
of $x_0$ in some scenarios.
 In addition the predictions we have
for all the mixings using these fitted parameters are also very good.

We conclude that the vector meson masses and mixings can be well described
within CHPT with the standard quark mass ratios.

\section*{Note added}
After this paper was submitted we became aware of a paper where the
value of $g$ has been determined from the rates of $\tau\to\omega\pi\nu$.
The values obtained there, \rcite{wise}, are in good qualitative agreement
with those used here, they correspond to $g\approx0.29$ and $g\approx0.32$
depending on the number of experimental bins included.

\section*{Acknowledgment}
P. G acknowledges a grant form the Spanish Ministry for
Education and Culture. P. T was granted the EU TMR program
under contract ERB 4001GT952585. He thanks the NORDITA and
The Niels Bohr Institute members for the warm hospitality
during the stay. 

\appendix
\section{Integrals}
\rlabel{integrals}
We quote here the integrals we use.
First the simple scalar ones
\be
i\mu_m(i m^2\mu_m)=\int \frac{d^dq}{(2\pi)^2}\frac{1(q^2)}{q^2 - m^2} =
\frac{i m^2(m^4)}{16\pi^2}
\left[\lambda-\log\left(\frac{m^2}{\mu^2}\right)\right]\,,
\ee
with $\lambda = 1/\epsilon-\gamma+\log(4\pi)+1$ and $d=4-2\epsilon$.
The others can all be expressed in terms of $iJ_0$.
\ba
iJ_0 &=&
\int\frac{d^dq}{(2\pi)^2}
\frac{1}{v\cdot q - \omega + i\eta}\frac{1}{q^2-m^2+i\eta}
\nonumber\\
&=&\frac{i}{16\pi^2}\left[-2\omega\lambda +
\omega\left[-2+2\log\left(\frac{m^2}{\mu^2}\right)\right]
-4(m^2-\omega^2) F(\omega,m)\right]\,.
\ea
The function $F(\omega,m)$ is
\ba
F(\omega,m) &=&\frac{1}{2\sqrt{\omega^2-m^2}}\log
\left(\frac{\omega+\sqrt{\omega^2-m^2}}{\omega-\sqrt{\omega^2-m^2}}\right)
\qquad(\omega > m)
\nonumber\\
&=&
\frac{1}{\sqrt{m^2-\omega^2}}
\left(\frac{\pi}{2}-{\rm arctg}\frac{\omega}{\sqrt{m^2-\omega^2}}\right)
\qquad(-m<\omega<m)
\nonumber\\
&=&\frac{1}{2\sqrt{\omega^2-m^2}}
\left\{\log\left(\frac{-\omega-\sqrt{\omega^2-m^2}}
{-\omega+\sqrt{\omega^2-m^2}}
\right) + 2i\pi\right\}\qquad(\omega < -m)\,.
\ea
We now define
\ba
\int\frac{d^dq}{(2\pi)^2}
\frac{\left[q_\mu ; q_\mu q_\nu; q_\mu q_\nu q_\alpha; q_\mu q_\nu q^2\right]}
{v\cdot q - \omega + i\eta}\frac{1}{q^2-m^2+i\eta}
&=&
\nonumber\\ &&\hskip-7cm i \left[
J_1 v_\mu ; J_2 g_{\mu\nu} + J_3 v_\mu v_\nu ;
J_4 (g_{\mu\nu}v_\alpha+g_{\mu\alpha}v_\nu+g_{\nu\alpha}v_\mu)
+J_5 v_\mu v_\nu v_\alpha ; J_6 g_{\mu\nu} + J_7 v_\mu v_\nu
\right]\,.
\ea
These can all be expressed in terms of $J_0$ and $\mu_m$:
\ba
J_1 &=& \omega J_0 + \mu_m
\\
J_2 &=& \frac{1}{d-1}\left((m^2-\omega^2)J_0-\omega \mu_m\right)
\\
J_3 &=& \omega J_1 - J_2
\\
J_4 &=& \omega J_2 +\frac{m^2}{d}\mu_m
\\
J_5 &=& \omega J_3 - 2 J_4
\\
J_6 &=& m^2 J_2
\\
J_7 &=& m^2 J_3\,.
\ea
The expressions for those with higher powers of the ``heavy'' propagator can
be derived by taking derivatives of the above expressions with respect
to $\omega$.
The only difficulty there is the derivative of the function $F$.
This is given by
\be
\frac{\partial F(\omega,m)}{\partial\omega} =
\frac{1}{m^2-\omega^2}\left(\omega F(\omega,m)-1\right)\,.
\ee

\section{Approximate Analytical Expressions}
\rlabel{formulas}

In this appendix we collect the explicit formulae for the masses,
 in the approximation of non-diagonal fields inside the sunrise
diagrams.\\
For the tree level contribution we find:
\begin{eqnarray}
\delta \rho^+ &=&
4 a_1 B_0 (m_u+m_d)+a_2 p_{ext}^2
+16 B_0^2 m_u m_d c_1+16 B_0^2 (m_u^2+m_d^2) c_2
- \delta_2+\frac{5}{9}  \delta_3\nonumber\\
\delta \rho^0 &=&
4 a_1 B_0 (m_u+m_d)+a_2 p_{ext}^2
+8 B_0^2 (m_u^2+m_d^2) (c_1+2c_2)+\frac{\delta_1}{2}
+\frac{5}{9}  \delta_3\nonumber\\
\delta \rho^0 \omega &=&
4 a_1 B_0 (m_u-m_d)+ 8 B_0^2 (m_u^2-m_d^2) (c_1+2 c_2)
+\frac{\delta_1}{6}+  \frac{\delta_3}{3}\nonumber\\
\delta \rho^0 \phi &=&
- \frac{\delta_1}{3\sqrt{2}}\nonumber\\
\delta \omega &=&
4 a_1 B_0 (m_u+m_d)+a_2 p_{ext}^2
+8 B_0^2 (m_u^2+m_d^2) (c_1+2 c_2)+ \frac{\delta_1}{18}
+\frac{5}{9}  \delta_3\nonumber\\
\delta \omega \phi &=&
- \frac{\delta_1}{9\sqrt{2}}\nonumber\\
\delta \phi &=&
8 a_1 B_0 m_s+a_2 p_{ext}^2
+16 B_0^2 m_s^2 (c_1+2 c_2)+ \frac{\delta_1}{9}+ 
\frac{2}{9} \delta_3
\nonumber\\
\delta K^{*+} &=&
4 a_1 B_0 (m_u+m_s)+a_2 p_{ext}^2
+16 B_0^2 m_u m_s c_1 +16 B_0^2 (m_u^2+m_s^2) c_2
- \delta_2+\frac{5}{9} \delta_3\nonumber\\
\delta K^{*0} &=&
4 a_1 B_0 (m_d+m_s)+a_2 p_{ext}^2
+16 B_0^2 m_d m_s c_1 +16 B_0^2 (m_d^2+m_s^2) c_2
+\frac{2}{9} \delta_3\nonumber
\end{eqnarray}
For the tad-pole type diagrams, we find:
\begin{eqnarray}
\delta \rho^+ &=&
a_1 B_0 \frac{2}{ F^2} \bigg(\mu_{\pi^0}(
m_u c_m^2 + m_d c_p^2 )
+2\mu_{\pi^+} (m_u+ m_d)
+\mu_{K^+}(m_u+m_s)+\mu_{K^0}(m_d+m_s) \nonumber\\&+&
 \mu_\eta ( m_u s_p^2 + m_d s_m^2 )\bigg) 
-\frac{2 b_1}{F^2}\bigg( \Sigma _{K^+} + \Sigma _{K^0} 
+2 \Sigma _{\pi ^+} + (c^2 + {s^2 \over 3})\Sigma _{\pi ^0}
+(s^2 + {c^2 \over 3})\Sigma _{\eta} \bigg) \nonumber\\
&+& \frac{2 b_2}{ F^2}\bigg( 
(s^2 - {c^2 \over 3}) \Theta _\eta +
(c^2 -{s^2 \over 3} ) \Theta _{\pi ^0} \bigg) \nonumber\\
\delta \rho^0 &=&
a_1 B_0 \frac{2}{ F^2} \bigg(\mu_{\pi^0}(
m_u  c_m^2
+ m_d  c_p^2 )
+2 \mu_{\pi^+}(m_u+m_d) +\mu_{K^+}(m_u+m_s)+\mu_{K^0}(m_d+m_s) 
\nonumber \\ &+&
 \mu_\eta ( m_u s_p^2 +
m_d s_m^2
)\bigg) 
-\frac{2 b_1}{ F^2}\bigg( \Sigma _{K^+} + \Sigma _{K^0} 
+2 \Sigma _{\pi ^+} + (c^2 + {s^2 \over 3})\Sigma _{\pi ^0}
+(s^2 + {c^2 \over 3})\Sigma _{\eta} \bigg) \nonumber\\
&+& \frac{2 b_2}{F^2}\bigg( 
2 \Theta _{\pi ^+}
-(s^2 + {c^2 \over 3}) \Theta _\eta 
-(c^2 +{s^2 \over 3} ) \Theta _{\pi ^0} \bigg) \nonumber\\
\delta \rho^0 \omega &=&
a_1 B_0 \frac{2}{ F^2} \bigg(\mu_{\pi^0}(
m_u c_m^2 - m_d c_p^2
)
+\mu_{K^+}(m_u+m_s)-\mu_{K^0}(m_d+m_s)
+ \mu_\eta ( m_u s_p^2
\nonumber\\&-&  m_ds_m^2 )\bigg) 
+\frac{2 b_1}{F^2} \bigg( \Sigma _{K^0} -\Sigma _{K^+} 
+ {2 c s \over \sqrt{3} } ( \Sigma _{\pi ^0} - \Sigma _{\eta } )
\bigg) 
+ \frac{4 b_2}{ \sqrt{3}  F^2} s c\bigg( 
\Theta _{\pi ^0} - \Theta _{ \eta} \bigg) \nonumber\\
\delta \rho^0 \phi &=& 
\frac{2 \sqrt{2} b_2}{ F^2} \bigg( 
\Theta _{K^0} - \Theta _{K^+} \bigg) \nonumber \\
\delta \omega &=&
a_1 B_0 \frac{2}{ F^2}\bigg(\mu_{\pi^0} ( 
m_u c_m^2 + m_d c_p^2)
+2\mu_{\pi^+} (m_u+m_d)
+\mu_{K^+}(m_u+m_s)+\mu_{K^0}(m_d+m_s)
\nonumber\\&+&
\mu_\eta (m_u s_p^2 + m_d s_m^2 ) \bigg) 
-\frac{2 b_1}{ F^2}\bigg( \Sigma _{K^+} + \Sigma _{K^0} 
+2 \Sigma _{\pi ^+} + (c^2 + {s^2 \over 3})\Sigma _{\pi ^0}
+(s^2 + {c^2 \over 3})\Sigma _{\eta} \bigg) \nonumber\\
&-& \frac{2 b_2}{ F^2}\bigg( 
2 \Theta _{\pi ^+}
+(s^2 + {c^2 \over 3}) \Theta _\eta 
+(c^2 +{s^2 \over 3} ) \Theta _{\pi ^0} \bigg) \nonumber\\
\delta \omega \phi &=& 
-\frac{2 \sqrt{2} b_2}{ F^2} \bigg( 
\Theta _{K^0} + \Theta _{K^+} \bigg) \nonumber \\
\delta \phi &=&
a_1 B_0 \frac{4}{ F^2}
\bigg(\frac{4}{3}\mu_{\pi^0} m_s s^2
+\mu_{K^+} (m_u+m_s)+\mu_{K^0}(m_d+m_s) +\frac{4}{3}
\mu_\eta m_s c^2
\bigg) \nonumber\\&-&
\frac{4 b_1}{F^2} \bigg(
\Sigma _{ K^+} +\Sigma _{ K^0} +{ 2 s^2 \over 3} \Sigma _{ \pi ^0}
+{ 2 c^2 \over 3} \Sigma _{ \eta} \bigg)
-\frac{8 b_2}{3 F^2} \bigg( 
c ^2\Theta _\eta + s^2 \Theta _{\pi ^0} \bigg) \nonumber  \\
\delta K^{*+} &=&
a_1 B_0\frac{2}{ F^2}\bigg( \mu_{\pi^0}(m_u
c_p^2 +\frac{4}{3}m_s s^2)+\mu_{\pi^+}(m_u+m_d) 
+2\mu_{K^+}(m_u+m_s)+\mu_{K^0}(m_d+m_s)
\nonumber \\ &+& 
\mu_\eta(m_u s_p^2
+\frac{4}{3}m_s c^2 )\bigg) 
-\frac{b_1}{F^2}\bigg( 4 \Sigma _{K^+}  +
2 \Sigma _{K^0} + 2\Sigma _{\pi ^+} +
( {5 c^2 \over 3} + {2 sc \over \sqrt{3} } +s^2) \Sigma _{\eta}
 \nonumber \\
&+&
( {5 s^2 \over 3} - {2 sc \over \sqrt{3} } +c^2) \Sigma _{\pi ^0} \bigg)
+ \frac{4 b_2}{ \sqrt{3} F^2} \bigg(
({c^2 \over \sqrt{3} } + sc ) \Theta _{\eta} +
({s^2 \over \sqrt{3} } - sc ) \Theta _{\pi ^0} \bigg) \nonumber \\
\delta K^{*0} &=&
a_1 B_0\frac{2}{ F^2}\bigg( \mu_{\pi^0}(m_d
c_p^2 +\frac{4}{3}m_s s^2) +\mu_{\pi^+}(m_u+m_d)
+\mu_{K^+}(m_u+m_s)+2\mu_{K^0}(m_d+m_s)
\nonumber \\ &+&
\mu_\eta(m_d  s_m^2 +\frac{4}{3}m_s c^2 )\bigg) 
-\frac{b_1}{F^2}\bigg( 2 \Sigma _{K^+}  +
4 \Sigma _{K^0} +  2 \Sigma _{\pi ^+} +
( {5 c^2 \over 3} - {2 sc \over \sqrt{3} } +s^2) \Sigma _{\eta} 
\nonumber \\ &+&
( {5 s^2 \over 3} + {2 sc \over \sqrt{3} } +c^2) \Sigma _{\pi ^0} \bigg)
+ \frac{4 b_2}{ \sqrt{3}F^2} \bigg(
({c^2 \over \sqrt{3} } - sc ) \Theta _{\eta} +
({s^2 \over \sqrt{3} } + sc ) \Theta _{\pi ^0} \bigg)
\end{eqnarray}

We have introduced the following notation:

\be
b_1 = a_5 + {a_7 + a_8 + a_{11}\over 4}  \qquad
b_2 = {a_6 \over 2 } + {a_9\over 4} + {a_{10}\over 8}
\ee
and
\ba
\mu _P &=& -{m_P^2 \over 16 \pi^2 } \log { m_P ^2 \over \mu ^2 } \qquad
\Sigma _P = m_P^2 \left(\mu _P + {m_P^2 \over 16 \pi ^2}
{ a_7 +a_8 + a_{11} \over 8 a_5 +2 (a_7 + a_8 + a_{11}) } \right) 
\nonumber \\
\Theta _P &=& m_P^2 \left(\mu _P + { m_P^2 \over 16 \pi ^2}
{ 2 a_9 +a_{10} \over 8 a_6 +4 a_9 + 2a_{10} } \right) 
\ea
We have also defined :
\ba
c&=& \cos \theta , \quad s=\sin \theta , \quad
c_p = c + {s\over \sqrt{3}} , \quad c_m = c - {s\over \sqrt{3} },
s_p = s+{c\over \sqrt{3}} ,\quad s_m = s -{c\over \sqrt{3} },
\ea

For the sunrise type diagrams, we find:

\ba
\delta \rho^+ &=& \Lambda \left\{  
{2\over 3} c^2 [\eta, \rho^+] + [k^+, K^{*0}] +  [k^0, K^{*0}] +
 2[\pi^+, \omega] + {2 \over 3} s^2 [\pi^0, \rho^+]
\right\} \nonumber \\
\delta\rho^0 &=& \Lambda \left\{ 2 \ s^2 \ [\eta,\omega] +
 {2\over 3} c^2 \ [\eta, \rho^0] + [k^+, K^{*+}] + [k^0, K^{*0}] + 
 2\ c^2\ [\pi^0 , \omega] + {2 \over 3} s^2\ 
 [\pi^0, \rho^0] \right\} \nonumber \\
\delta \rho^0 \omega &=& \Lambda \left\{
{2 \over \sqrt{3}} cs [\eta, \omega] + 
{2 \over \sqrt{3}} cs [\eta, \rho^0] + [k^+, K^{*+}] - 
[k^0, K^{*0}] - {2 \over \sqrt{3}} cs [\pi^0, \omega] - 
{2 \over \sqrt{3}}   cs [\pi^0, \rho^0] 
\right\} \nonumber\\ 
\delta \rho^0 \phi &=& \sqrt{2}  \Lambda \left\{
[k^+, K^{*+}] - [k^0, K^{*0}]
\right\} \nonumber \\
\delta \omega &=& \Lambda \left\{
 {2 \over 3} c^2 [\eta, \omega] + 
   2 s^2 [\eta, \rho^0] + [k^+, K^{*+}] + [k^0, K^{*0}] + 
   4 [\pi^+, \rho^+] + {2 \over 3} s^2 [\pi^0, \omega ] + 
   2 c^2 [\pi^0, \rho^0] \right\} \nonumber \\
\delta \omega  \phi  &=& \sqrt{2}  \Lambda \left\{ 
[k^+, K^{*+}] + [k^0, K^{*0}]
\right\} \nonumber \\
\delta\phi &=& \Lambda \left\{
{8 \over 3} c^2 [\eta, \phi] + 2 [k^+, K^{*+}] + 
   2 [k^0, K^{*0}] + {8 \over 3} s^2 [\pi^0, \phi]
\right\} \nonumber \\
\delta K^{*+}&=& \Lambda \left\{
{1 \over 6}  c^2 [\eta, K^{*+}] - 
{cs \over \sqrt{3} } [\eta, K^{*+}] + {s^2\over 2} [\eta, K^{*+}] + 
{1\over 2} [k^+, \omega] + [k^+, \phi] +{1\over 2} [k^+, \rho^0] + 
[k^0, \rho^+] \right. \nonumber \\ &+& \left. [\pi^+, K^{*0}]  
+  {c^2 \over 2} [\pi^0, K^{*+}] + 
{cs \over \sqrt{3}} [\pi^0, K^{*+}] + {s^2 \over 6} [\pi^0, K^{*+}]
\nonumber \right\} \\
\delta K^{*0}&=& \Lambda \left\{
{c^2 \over 6 } [\eta, K^{*0}] + 
  {cs \over \sqrt{3} } [\eta, K^{*0}] + {s^2 \over 2} [\eta, K^{*0}] + 
 [k^+, \rho^+] + {1\over 2}[k^0, \omega] + [k^0, \phi]\right.
\nonumber\\& +& 
{1\over 2}[k^0, \rho^0] + [\pi^+, K^{*+}] 
+ \left.
 {c^2 \over 2}[\pi_0, K^{*0}] -{cs \over \sqrt{3}}[\pi^0, K^{*0}] +
   {s^2 \over 6} [\pi^0, K^{*0}] \right\} \nonumber
\ea

We have introduced 

\be
\Lambda = {4 g^2 \over F^2  }, \qquad
[G,\Omega] = \left( 1+{p\cdot v \over m_V} -{m_G^2 \over 2 m_V}
{\partial \over \partial m_\Omega } \right)
J_2(m_G,m_\Omega -m_V - {p\cdot v} ) \nonumber\\
\ee

Where the $[G,\Omega]$ function is related with the one defined in
\rref{vectordiag}.
The physical  $\eta_F$ and $\pi_{0}$ are related to the isospin
eigenfields $\eta_8$ and $\pi_0 ^I$ by

\be
\eta_8 = \cos \theta \ \eta_F - \sin \theta \ \pi_{0} \qquad
\pi_0 ^I  = \sin \theta \ \eta_F + \cos \theta \ \pi_{0} , \nonumber\\
\ee

\end{document}